# Structural phase transition of ternary high-k dielectric SmGdO$_3$: Evidence from ADXRD and Raman Spectroscopic Studies

Yogesh Sharma[1], Pankaj Misra[1], Satyaprakash Sahoo[1], A. K. Mishra[2], S. M. Sharma[2], and Ram S. Katiyar[1]

[1]Department of Physics and Institute for Functional Nanomaterials, University of Puerto Rico, San Juan, PR 00936-8377, USA

[2]High Pressure & Synchrotron Radiation Physics Division, Bhabha Atomic Research Centre, Mumbai

## Abstract

High-pressure synchrotron based angle dispersive x-ray diffraction (*ADXRD*) studies were carried out on SmGdO$_3$ (SGO) up to 25.7GPa at room temperature. *ADXRD* results indicated a reversible pressure-induced phase transition from ambient monoclinic to hexagonal phase at ~ 8.9 GPa. The observed pressure-volume data were fitted into the third–order Birch-Murnaghan equation of state yielding zero pressure bulk moduli B$_0$ = 132(22) and 177(22) GPa for monoclinic and hexagonal phases, respectively. Pressure dependent micro-Raman spectroscopy further confirmed the phase transition. The mode Grüneisen parameters and pressure coefficients for different Raman modes corresponding to each individual phase of SGO were calculated.

Corresponding email: yks181086@gmail.com, satya504@gmail.com, rkatiyar@hpcf.upr.edu

# Introduction

The rare-earth sesquioxides $RE_2O_3$, (where RE = rare-earths) have attracted great deal of interest from a scientific and technological viewpoint.[1-6] Due to their various polymorphs and variety of applications, such as laser rods, phosphors, abrasive and refractory materials. The partial filling of the inner 4f electron shells, leading to well known lanthanide contraction, affects the physical properties of rare earth metals as well as rare earth sesquioxides.[3-6] However, 4f electrons do not participate in bonding and behave like core electrons. The screening of the 4f electrons by the Xe core causes the bulk properties of lanthanides relatively unchanged with addition of electrons in the 4f shell. That is why the properties of the trivalent lanthanides vary in a gradual manner as one moves towards the end member of the series. Furthermore, the exotic and rich physical properties of these oxides could be tuned by the external pressure and temperature which induce significant crystallographic and electronic changes.[5-7]

Recently, some of the rare-earth sesquioxides (REOs) have attracted considerable attention for application as high-k dielectric materials to replace conventional silicon dioxide in complementary metal-oxide semiconductor (CMOS) technology.[8-10] Interestingly, some of REOs were found to show variable dielectric constant depending on their crystal structure. For instant, $Pr_2O_3$ in its cubic phase shows dielectric constant of ~ 15 which increases to ~ 25 in the hexagonal phase.[11] This tunability of dielectric constant with structural effect could be important in thin film studies of these oxides for gate-dielectric applications. Some of the REOs like, $La_2O_3$, $Sm_2O_3$, and $Gd_2O_3$ with dielectric constant ~ 27, 15 and 13 respectively, have been studied as promising alternative dielectrics with a large optical band gap and good thermal stability. However, it has been observed that the moisture absorption is another problem with some of the REOs because of their high hygroscopic nature, which induces permittivity deterioration.[12] Therefore, the ABO$_3$-type ternary rare-earth oxides RE′RE″O$_3$ (where RE′ and RE″ are two different rare-earths) have been proposed to solve the problems of binary REOs. Ternary oxides LaGdO$_3$ (LGO)[13], LaLuO$_3$ (LLO)[14], and SmGdO$_3$ (SGO)[15] were reported to show better properties as high-k dielectric materials in comparison to their constituent REOs. Recently, we reported multilevel resistive memory switching in amorphous SGO thin film based Pt/SGO/Pt metal-insulator-metal structures for non-volatile resistive random access memory (ReRAM) applications.[16] In other applications, the solid solution of Sm$_{0.2}$Gd$_{0.8}$O$_3$ has also been reported as potential material for fluorescence labelling.[17]

In comparison to the binary REOs, there are only a few studies on structural properties of ternary REOs, especially under high pressure. Herein, we present structural phase transition studies of ternary SGO using *in-situ* angle dispersive x-ray diffraction (*ADXRD*) and micro-Raman scattering where the ambient monoclinic (B type) phase of SGO was found to transform into high pressure hexagonal (A type) phase. In what follows, we provide a detailed pressure dependent structural evolution of SGO based on structural refinement of *XRD* data, where pressure-volume data are fitted into the third − order Birch-Murnaghan equation to obtain the bulk moduli of the B and A phases. The mode Grüneisen parameters for different Raman modes of each individual phase were also calculated from the pressure dependent Raman analysis.

## Synthesis

Polycrystalline powder samples were prepared by solid state reaction method by taking the proper stoichiometric quantities of high purity Sm$_2$O$_3$ and Gd$_2$O$_3$. These stoichiometric powders were mixed thoroughly, placed into Al$_2$O$_3$ crucibles and then fired in air at 1250°C for 12 h. The resultant powders were ground and re-sintered at 1400°C for 15 h. Phase purity of the as synthesized SGO powder was confirmed by Rietveld refinement of the X-ray powder diffraction data taken in the 2θ range of 20-80° at ambient conditions using Rigaku ultima III X-ray diffractometer.

**Experimental Details**

The *in-situ* high pressure *ADXRD* experiments were performed at beam line BL11[18] at INDUS 2 synchrotorn source in India. The monochromatic x-ray of wavelength of 0.6199 Å is employed to carry out *ADXRD* studies. The modified Mao-Bell type of diamond anvil cell with a culet size of 0.4 mm was used for generating pressure up to ~ 25.7 GPa. The Image plate (IP) kept at a distance of ~ 20 cm from the diamond anvil cell was used for detection of diffracted x-rays. The sample to detector distance was calibrated using $CeO_2$. The powdered sample (SGO) along with a few specs of copper (Cu), used as a pressure marker, was loaded inside a sample chamber, diameter 100 μm, made up of a tungsten gasket pre-indented to a thickness of 60 μm. High pressure Raman measurements were performed in a Mao-Bell type Diamond anvil cell (DAC) with a culet size of 0.6mm for the diamond anvils. A stainless-steel gasket with the thickness of 50μm and 100μm hole drilled in the center has been used as a sample chamber. A limited amount of powder sample was loaded in central hole together with the methanol/ethanol (4:1) mixture as a pressure transmitting medium. For pressure calibration, a ruby crystal was put in the sample chamber hole and the pressure value was measured by applying the standard ruby fluorescence method.[19] The micro-Raman study was performed in the backscattering geometry by using a Jobin-Yvon T64000 triple spectrometer with grating (1800 grooves $mm^{-1}$). The 532 nm line of a diode pumped solid state laser (Innova 90-5) was used as excitation source.

**Result and Discussion**

The X-ray diffraction pattern (outside the DAC) of as synthesized SGO powder at ambient temperature and pressure is shown in Fig. 1 (a). The Rietveld refinement of the diffraction pattern was performed by taking monoclinic (B-type) C2/m symmetry. The refined results shown in Fig. 1(a) demonstrated excellent fitting without change in the peak profile or relative intensity, confirming pure monoclinic phase thus ruling out the existence of mixed phase (cubic + monoclinic) formation as observed in case of constituents REOs at ambient conditions.[1,2] Here it is noteworthy that the constituent REOs crystallise into cubic phase or in a mixed phase (cubic+monoclinic) while this ternary REO (SGO) crystallises into monoclinic structure at ambient conditions. Thus we can speculate that the effect of cationic radii of rare earths Sm and Gd resembles with that of the high pressure effect where some cubic REOs undergo structural phase transition to monoclinic phase. This can be further understood with the fact that the chemical pressure exerted by the Rare earth cations in ternary REOs causes these sesquioxide to stabilise the high pressure monoclinic phase at ambient conditions. Further, the refined unit cell parameters ($a$ = 14.1377 Å, $b$ = 3.601478 Å, $c$ = 8.8117 Å, and $\alpha$ = 90°, $\beta$ = 100.109°, $\gamma$ = 90°) were found to be in better agreement with those reported for isostructural LGO at ambient conditions.[20] Using the obtained unit cell parameters and atomic positions the crystal structure of monoclinic SGO was generated as shown in Fig.1 (b). Figure 2(a) shows the pressure dependent angle resolved x-ray diffraction patterns of SGO upon compression as well as decompression. At pressure values of < 7GPa only monoclinic phase was found to exist. The refined *ADXRD* pattern of SGO at 0.4GPa again confirmed the single monoclinic phase, as shown in Fig. 2(b). Broadening of the diffraction peaks along with the appearance of some new peaks was observed when increasing the pressure above 7.1GPa. Both hexagonal and monoclinic (A+ B) phases were found to coexist between the pressure range of 7.1- 8.9GPa, which was confirmed by the mixed phase refinement of *ADXRD* pattern, as shown in Fig. 2(c). Above 8.9GPa, the diffraction peaks corresponding to B-phase disappeared, accompanied by the significant increase of A-phase peak intensities, indicating a complete phase transition from mixed phase (A+B) to pure heaxagonal A-phase. This high pressure A-phase was further confirmed by refining the *ADXRD* pattern based on hexagonal A-type structure with space group of P $\bar{3}$m1, as depicted in Fig. 2(d), where the calculated *ADXRD* pattern was found to fit the experimentally observed pattern nicely. On further raising the pressure up to 25.7 GPa, the x-ray diffraction peaks of hexagonal phase move towards higher two theta values. This implies the stability of A-phase up to the highest pressure. Therefore, our studies confirmed that the B → A+B and A+B → A phase transition in SGO occurred at ~ 7.1 and 8.9GPa respectively. The pressure value for the complete transformation from B → A was found to be lower than the constituent oxides $Sm_2O_3$ and $Gd_2O_3$, where B → A phase transformation were reported to be around 10.8 and 12 GPa respectively.[1,2] Also, one of the interesting observations which could be different from the sesquioxides is that the absence of cubic (C-type) and/or mixed (cubic + monoclinic) phase in ternary SGO at ambient as well as intermediate pressure range. From the earlier theoretical and experimental reports, the sesquioxides $Sm_2O_3$ and $Gd_2O_3$ were found to be in mixed (cubic + monoclinic ) phase at ambient conditions.[1,2] In these sesquioxides monoclinic phase act as metastable state which converts to hexagonal phase at normal pressures, where as the cubic phase transforms to the hexagonal phase at higher pressures.[1] In ternary SGO, only monoclinic phase was found to exist at ambient conditions and we did not found any signature of cubic phase throughout the pressure range. Further, as depicted in Fig. 2(a), the *ADXRD* patterns upon decompression process also showed the transformation of hexagonal phase to monoclinic phase when the pressure was released, which confirms the reversibility of the B → A phase transition.

Moreover, the pressure dependence of unit-cell volume of the monoclinic and hexagonal phases were fitted into the third-order Birch-Murnaghan (B-M) equation of state:[21]

$$P(V) = \frac{3B_0}{2}\left[\left(\frac{V_0}{V}\right)^{7/3} - \left(\frac{V_0}{V}\right)^{5/3}\right] \times \left[1 + \frac{3}{4}(B_0' - 4)\left\{\left(\frac{V_0}{V}\right)^{2/3} - 1\right\}\right] \qquad (1)$$

where, $B_0$ and $B_0'$ represent the bulk modulus and its pressure derivative at zero pressure, respectively. $V$ and $V_0$ correspond to the unit-cell volumes at pressure P and atmospheric pressure, respectively. Figure 3 shows the pressure induced variation of volume fitted with the third order Birch Murnaghan equation of state. The symbols represent the observed P-V data while the solid line is obtained from B-M fit of the observed data. The bulk modulus are determined to be 132(22) and 177(9) GPa for monoclinic and hexagonal phases, respectively. The values of bulk modulus of B and A-phase in SGO revealed that the high pressure hexagonal phase becomes less compressible than the ambient phase.[1]

The above results regarding structural phase transition of SGO was also supported by pressure dependent Raman scattering studies. As per group theory analysis, the monoclinic SGO with C2/m (No. 12) space group should show 21 Raman-active modes at the Brillouin zone center: $\Gamma = 14A_g + 7B_g$, and there are only four Raman-active modes corresponding to high pressure hexagonal phase with space group P $\bar{3}$m1 (No. 224): $\Gamma = 2A_{1g} + 2E_g$.[20,22] Raman spectra recorded at a few representative pressure during compression as well as on release have been depicted in Fig. 4(a), where 15 out of 21 Raman modes are distinguished corresponding to ambient monoclinic phase below 2.98 GPa. With increasing pressure, the modes at higher frequencies started to disappear. On further raising the pressure beyond 2.98 GPa, two new modes at ~ 193 and ~ 456 cm$^{-1}$ were found to emerge indicating the appearance of new phase. At ~ 8GPa, only four Raman peaks were observed, which can be indexed as the two bending ($A_{1g}$) and two stretching ($E_g$) modes of the A-type hexagonal structure based on the previous studies on A-type sesquioxides.[2,23] It is also evident from fig.

4(a) that after complete release of pressure the Raman modes corresponding to ambient monoclinic (B type) phase reappears implying the reversibility of the pressure induced phase transition in SGO. It is important to note that we could not found any signature Raman peak corresponding to cubic phase hence existence of mixed (cubic + monoclinic) can be ruled out, whereas in $Sm_2O_3$ and $Gd_2O_3$ the most intense Raman peak has been reported to be assigned as combination of $A_g$ and $F_g$ modes of the cubic phase at ambient conditions along with peaks corresponding to monoclinic phase.[1,2] Thus the results obtained by Raman studies are consistent with those observed by *ADXRD* measurements, where a complete transition from monoclinic to hexagonal phase occurred at ~8 GPa, slightly lower than the 8.9GPa obtained from *ADXRD* measurements. However, B → A phase transition was found to start at ~ 2.98GPa, which is quite lower than the pressure value of 7.1GPa as observed from *ADXRD*. This type of phenomenon of the difference in the transition pressure calculated using *XRD* and Raman scattering studies was reported in $Sm_2O_3$.[2,24] Recently, the Raman scattering studies by Jiang et al.[2] also reported lower value of the transition pressure for B → A and C → A phase transitions compared to the pressure values obtained from *ADXRD*. According to these previous studies,[2, 24] one possible reason for this lowering in transition pressure could be due to the extra-sensitivity of the Raman modes of the hexagonal phase towards changes in the chemical bonding while increasing the pressure, and therefore become reasonably distinctive at lower pressures during the phase transformation.

The pressure dependence of the phonon frequencies for the observed Raman modes of SGO was depicted in Fig. 4(b). The modes were observed to shift towards higher frequency, possibly due to contraction in Sm-O and/or Gd-O bonds under high pressures,[1,2] where the absence of modes softening on compression indicated that the B → A transition is first order.

The values of mode Grüneisen parameters were determined for each individual phase of SGO using the relation:

$$\gamma = \left(\frac{B_0}{\omega_{0i}}\right) \cdot \left(\frac{d\omega_i}{dP}\right) \qquad (2)$$

where, $B_0$ is the bulk moduli (132 and 177 for B and A phases, respectively) $\omega_{0i}$ is the mode frequency at ambient pressure, and $d\omega_i/dP$ is their pressure derivative or pressure coefficient. The values of $\omega_{0i}$ for hexagonal phase were calculated using extrapolation of the experimental values to ambient pressure. Table I summarizes the values of mode frequencies $\omega_{0i}$, their pressure derivatives, and mode Grüneisen parameters for each individual phase of SGO.

## Conclusions

The structural phase transition in high-k dielectric material SmGdO$_3$ was revealed by high-pressure synchrotron *ADXRD* and micro-Raman scattering measurements. *ADXRD* studies under high pressure indicated a structural transition from monoclinic to hexagonal phase at ~ 8.9GPa. The hexagonal phase was found to be stable up to 25.7GPa and transforms to monoclinic phase after decompression to ambient conditions. This structural transition was further verified using Raman spectroscopy. The third order B-M equation fitting of pressure-volume data yielded zero pressure bulk moduli $B_0$ = 132(22) and 177(9) GPa for the monoclinic and hexagonal phases, respectively. Further, the mode Grüneisen parameters and pressure coefficients of different Raman modes for each individual phase of SGO were also determined.

**Acknowledgements:** Financial support from DOE Grant No. DE-FG02-08ER46526 is acknowledged. The authors (Y. S. and P. M.) are grateful to IFN for fellowship under NSF-RII-0701525 grant.

**Figure captions:**

Figure1 (a): The Rietveld refined powder XRD pattern of monoclinic SGO at ambient conditions outside the DAC. (b) Schematic unit cell structure of B-type monoclinic SGO.

Figure 2: (a) Diffraction pattern of SGO recorded inside the DAC were stacked at a few representative pressures. The XRD peaks from sample are marked with (hkl) while that from pressure marker and gasket are marked as Cu and G respectively. Letter (r) adjacent to pressure values shows the diffraction pattern on decompression. Rietveld refined diffraction pattern of SGO stacked at (b) 0.4 GPa, (c) 7.0 GPa and (d) 8.9 GPa. The solid circle symbols represent the observed diffraction pattern of $SmGdO_3$. The red, green and blue coloured solid lines represent the calculated diffraction pattern, fitted background and difference of observed and calculated diffraction pattern respectively. The violet and magenta coloured vertical bars indicate the x-ray diffraction peak positions of B and A phase of SGO, respectively, while the black and orange coloured vertical bars indicate the x-ray diffraction peak positions of pressure marker (copper) and gasket (W) respectively.

Figure 3: Experimentally observed pressure-volume data of SGO. Solid symbols correspond to compression (filled) and decompression (hollow) data of SGO while the solid line is obtained by fitting the observed P-V variation with third order B-M equation of state.

Figure 5: (a) Raman spectra of SGO with increasing pressure up to 8GPa and after fully release of pressure. The mode assignment has been done based on the previous studies on isostructural compound LGO (ref. 20 and 22). (b) Pressure dependence of Raman frequencies of SGO, where the different color contrast represent the pressure regions related to different structural transitions.

**Table:**

Table I: Mode frequencies, their pressure derivatives, and Grüneisen parameters for phonon modes in monoclinic and hexagonal phases of SGO.

| \multicolumn{4}{c|}{Monoclinic (B) SGO} | \multicolumn{4}{c}{Hexagonal (A) SGO} |
|---|---|---|---|---|---|---|---|
| Modes | $\omega_0$ | $d\omega/dP$ | $\gamma$ | Modes | $\omega_0$ | $d\omega/dP$ | $\gamma$ |
| $B_g$ | 70.00(7) | 0.82(9) | 1.54(6) | $E_g$ | 89.09(7) | 1.39(9) | 2.76(1) |
| $A_g$ | 81.28(2) | 1.35(9) | 2.19(2) | $A_{1g}$ | 177.41(9) | 3.83(1) | 3.82(1) |
| $B_g$ | 95.58(5) | 1.33(4) | 1.83(6) | $A_{1g}$ | 441.73(1) | 2.60(1) | 1.04(1) |
| $A_g$ | 107.92(9) | 2.88(8) | 3.52(2) | $E_g$ | 468.96(4) | 3.10(3) | 1.17(3) |
| $A_g$ | 173.96(3) | 3.55(6) | 2.69(3) | | | | |
| $A_g$ | 216.67(8) | 3.33(7) | 2.02(7) | | | | |
| $A_g$ | 246.85(3) | 2.48(8) | 1.32(6) | | | | |
| $A_g$ | 259.39(3) | 4.72(4) | 2.40(1) | | | | |
| $A_g$ | 289.56(8) | 6.38(6) | 2.90(8) | | | | |
| $A_g$ | 376.75(3) | 4.75(5) | 1.66(4) | | | | |
| $B_g$ | 405.37(1) | 4.92(5) | 1.60(2) | | | | |
| $B_g$ | 417.71(5) | 3.94(9) | 1.24(5) | | | | |
| $A_g$ | 432.01(9) | 10.26(7) | 3.13(4) | | | | |
| $A_g$ | 471.207 | 6.97(7) | 1.95(2) | | | | |
| $A_g$ | 581.52(2) | 4.78(4) | 1.08(5) | | | | |

**Figure 1 (a) and (b)**

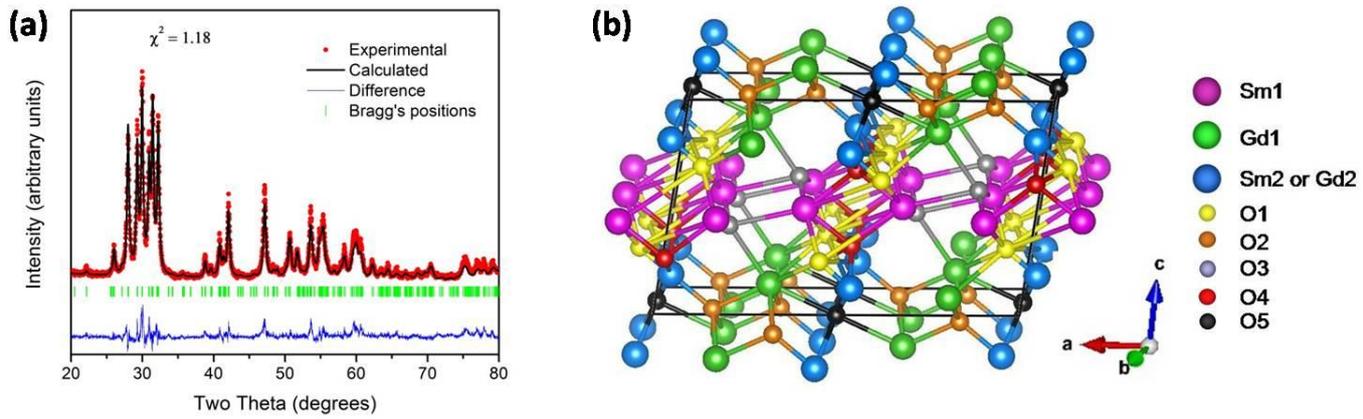

**Figure 2**

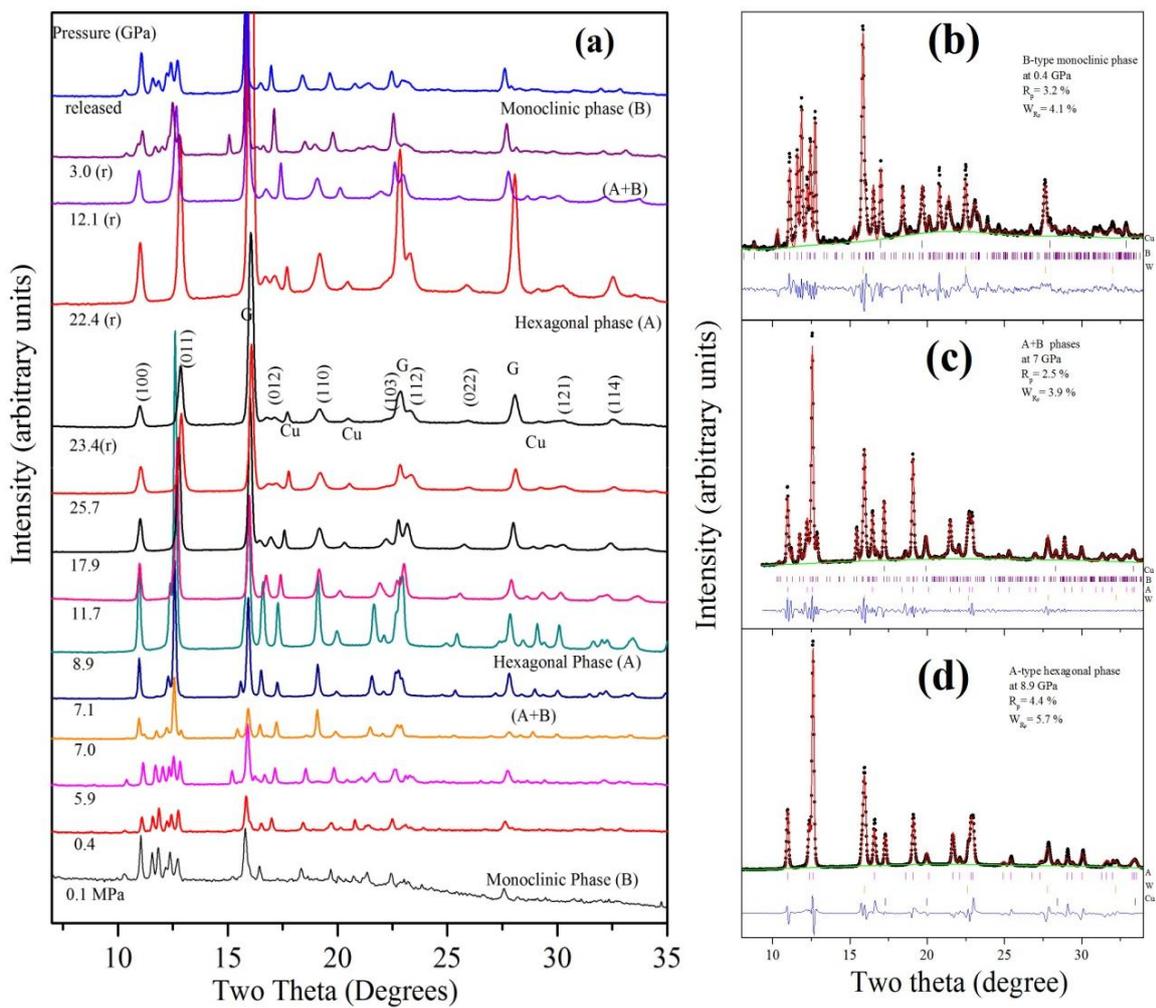

**Figure 3**

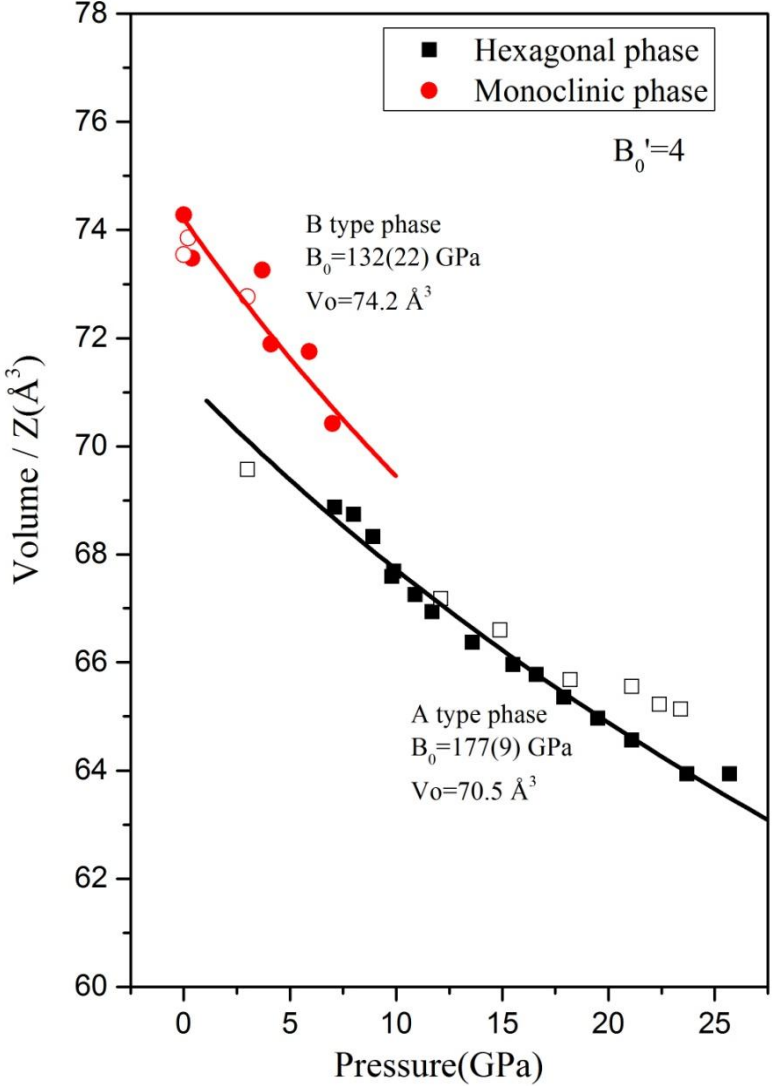

**Figure 4**

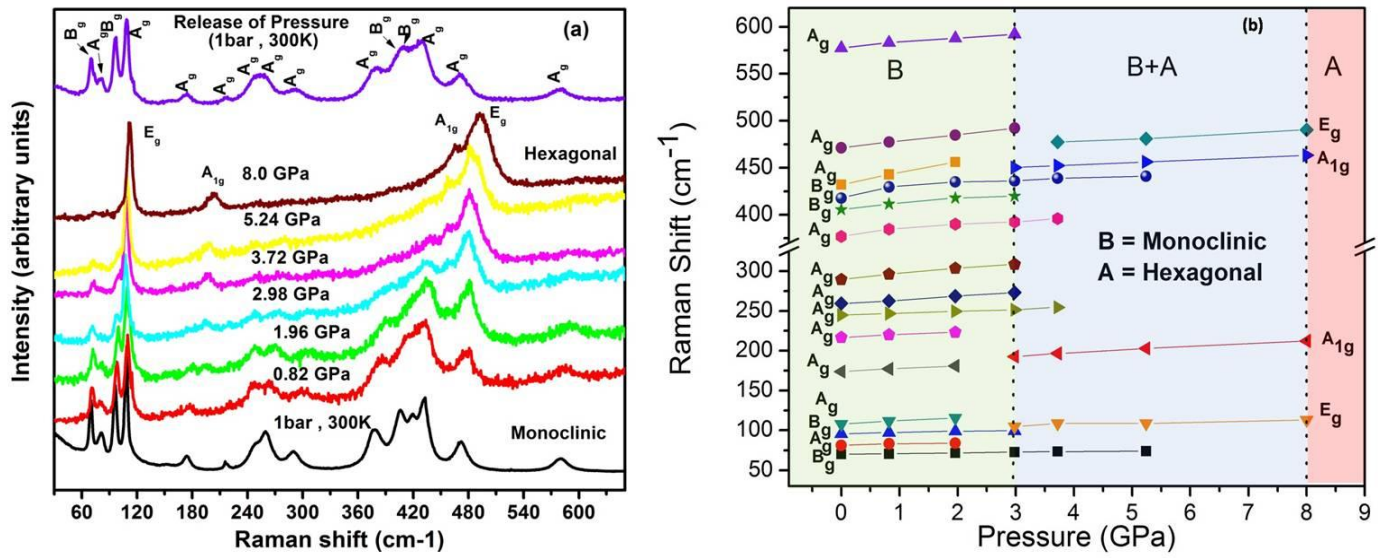